% this is a paper on the duration of grb's as a function of energy

\magnification = 1200
\raggedbottom

\def\ce{\centerline}
\def\ni{\noindent}
\def\lsim{\lower.5ex\hbox{$\;\buildrel < \over \sim \;$}}
\def\gsim{\lower.5ex\hbox{$\; \buildrel> \over \sim \;$}}
\def\ea{{\it et al.~}}
\def\eaa{{\it et al.}}
\def\ref{\noindent \hangafter=1 \hangindent=0.7truecm}

\def\cm{\ifmmode {\rm cm}^{-1} \else cm$^{-1}$ \fi}
\def\s{\ifmmode {\rm s}^{-1} \else s$^{-1}$ \fi}
\def\cc{\ifmmode {\rm cm}^{-3} \else cm$^{-3}$ \fi}
\def\cs{\ifmmode {\rm cm}^{-2} \else cm$^{-2}$ \fi}
\def\gr{$\gamma$-ray}

\def\g{\ifmmode \gamma \else $\gamma$\fi}
\def\gs{\ifmmode \gamma~ \else $\gamma~$\fi}
\def\ls{\ifmmode \lambda~ \else $\lambda~$\fi}
\def\l{\ifmmode \lambda \else $\lambda$\fi}
\def\G{\ifmmode \Gamma \else $\Gamma$\fi}
\def\ref{\noindent \hangafter=1 \hangindent=0.7 truecm}

\ce{\bf THE DURATION--PHOTON ENERGY RELATION OF GAMMA RAY}
\ce{\bf  BURSTS AND ITS INTERPRETATIONS}

\bigskip

\bigskip

\bigskip

\ce{Demosthenes Kazanas$^1$, Lev G. Titarchuk$^2$ and Xin--Min Hua$^3$}

\ce{Laboratory for High Energy Astrophysics}

\ce{Goddard Space Flight Center}

\ce{Greenbelt, MD 20771}

\vskip 3.0 truecm

\ce{\bf Abstract}

\bigskip

Fenimore \ea (1995) have recently presented a very tight
correlation between the spectral and the temporal structure in Gamma
Ray Bursts  (hereafter GRBs). In particular, they discovered that the
durations of the constituent subpulses which make up the time profile
of a given GRB have a well defined power--law dependence, of index
$\simeq 0.45$, on the energy $E$ of the observed photons. In this note we
present two simple models which can account in a straightforward 
fashion for the observed correlation. These models involve: (a) The 
impulsive injection of a population of relativistic electrons and
their subsequent cooling by synchrotron radiation. (b) The impulsive
injection of monoenergetic high energy photons in a medium of
Thompson depth $\tau_T \sim 5$ and their subsequent downgrading in
energy due to electron scattering. We present arguments for
distinguishing between these two models from the existing data. 

\vfill

\ni $^1$  email: kazanasl@heavx.gsfc.nasa.gov

\ni $^2$  also CSI Institute, George Mason University,  
4400 University Drive, Fairfax, VA 22030;

\ni ~~email: 
titarchuk@lheavx.gsfc.nasa.gov

\ni $^3$  NAS--NRC Senior Research Associate;~~
email:hua@rosserv.gsfc.nasa.gov

\eject

\baselineskip = 0.5truecm
\parskip = 4pt

\ce {\bf I. Introduction}

\bigskip

The issue of the nature of GRBs, a contentious one since their
discovery, has only deepened with the more recent observations by
BATSE  which confirmed their isotropic distribution in the sky while
indicating a significant lack of faint events, inconsistent with a
uniform distribution in space (Meegan et al 1992).  These facts have
provided support to models invoking the cosmological origin of the
GRBs, thus increasing further the uncertainty associated with the
location, nature and origin of the GRB phenomenon.  
This uncertainty is due
largely to the lack of association of any class of interesting
astrophysical objects within the positional error boxes of GRBs, which
would set the distance scale and hence the luminosity associated with
these events. In the absence of such positional association, attempts
to gain understanding on the nature of GRBs have relied in  searches 
for systematics  in the time and/or energy domain of the observed  
bursts. Part of our continuing ignorance with respect to the nature 
of GRbs is due to the absence of any significant 
correlations between the GRB spectral and/or temporal properties
(Paczynski  1992). 

This situation seems to  have recently changed:  Fenimore \ea
(1995) have  presented a very well defined correlation between the
average duration of the subpulses usually present in the time profile
of a given burst -- or the duration of the entire burst in the absence
of subpulses -- and the energy channel of observation. They have found
that the average pulse width, $\Delta \tau$, as determined either by the
autocorrelation function of the entire GRB or by specific fits of the
time profiles of individual subpulses in an given profile, is well
fitted by a power law in the energy $E$ at which the observation is
made; more specifically they have found that $\Delta \tau \simeq A
E^{-0.45 \pm 0.05}$, where $A$ is a proportionality constant. The above
reference thus made quantitative an effect that  had been noticed
earlier by Fishman \ea (1992) and Link, Epstein \& Priedhorsky
(1993) who indicated that, in general, the various peaks in the GRB
time profiles are shorter and better defined at higher energies. This
unique piece of information may, therefore,  serve as a stepping
stone in probing the physical processes associated with the radiation
emission in GRBs and maybe the entire GRB phenomenon.   
More recently, Norris \ea (1996) have further elaborated on the 
time profiles of GRBs as a function of the  energy channel of observation.
These authors have found that the relation reported by Fenimore \ea (1995) 
indeed holds but it may have slightly different slopes depending on the 
overall position of the observed pulse within the entire burst.

In this note we present two generic models which can 
provide an account of the
particular functional form of the observed correlation between
$\Delta \tau$ and $E$. It is of interest to note that the physical
processes associated with 
these two specific models are fundamentally different,
despite the fact that they both can  produce the observed pulse duration --
energy correlation. However, these models are quite simple in their
concept and specific enough to consider possible that further
observations and detail fits would yield systematics which could 
discriminate between them or will lead to further insights into the 
radiation mechanism of GRBs.  The first model involves the impulsive
injection of a power law of relativistic electrons in a given volume
and their subsequent cooling by synchrotron or inverse Compton
radiation, while the second one the impulsive injection of
high energy photons in an cold medium of Thompson depth $\tau_T \sim
10$ and their subsequent downgrading by electron scattering. 

In section II the mathematical formulation of both models is
presented along with sample time--dependent spectra and model pulse
durations at various frequencies. In section III the results are
discussed and conclusions are drawn concerning the impact of these
results on GRBs models.

\bigskip

\ce{\bf II. The Models}

\medskip

To begin with, it is worth noting that the relation discussed by  
Fenimore \ea (1995) precludes from the outset the possibility  
that the observed spectra are the result of up--Comptonization of soft 
photons by repeated scatterings in a steady--state, hot electron 
population (see e.g. Sunyaev \& Titarchuk 1980); in this case the 
the higher energy photons are those which would have spent longer time 
within the hot, scattering medium, leading to a correlation between the
burst duration $\Delta \tau$ and the photon energy $E$ of opposite 
sense to that observed. 

Before we discuss the models for the energy dependence of 
GRB duration in detail, we
present some simple, heuristic arguments which qualitatively indicate
that a well defined spectro -- temporal relation similar to that observed
is not indeed at all surprising. These considerations are 
motivated by the recent cosmological scenario for GRBs suggested by
M\'esz\'aros \& Rees (1993) and also by some arguments concerning 
photon scattering in a thick media and should generally be 
as applicable to galactic GRB models. It is worth noting here that, 
despite their apparent dissimilarity,
both models outlined below describe essentially the same process,
namely the energy degradation  of a population of relativistic 
particles, with  one of the models  refering to relativistic electrons 
while the other to photons. What is more important to bear in mind
is that the observed correlation appears to require the presence
of some sort of ``cooling"  in GRBs.

Consider the impulsive injection of relativistic particles at a given
volume and assume the emission to be optically thin. If synchrotron
or inverse Compton losses dominate the evolution of the electron
distribution, then the loss rate per electron is $\dot \g \propto B^2
\g^2$, where \gs is the Lorentz factor of a given electron and B the
magnetic field (if the losses are dominated by inverse Compton, then
$B^2$ should be replaced by the ambient photon energy density).
Therefore, the characteristic electron life time is $\Delta \tau \sim
\g  /\dot\g \propto \g^{-1}$. Since the characteristic energy of
synchrotron (or inverse Compton) emission is $E \propto \g^2 
\epsilon_0$ ($\epsilon_0$ is a characteristic energy), 
expressed in terms of $E$ rather than \g, the time duration is 
$\Delta \tau \propto E^{-1/2}$.

Consider, alternatively, the impulsive injection of monoenergetic
high energy photons of energy $E_0$ in a medium of Thomson depth
$\tau_T \gg 1$ and of electron temperature $kT \ll E_0$. Photons of
energy $E$ suffer fractional energy loss of order $\Delta E /E = 
(1 - \mu) E/m_ec^2$ per collision, where $\mu$ is the cosine of 
the photon scattering angle, or in terms of the photon wavelength 
$\lambda$, $\Delta \lambda = (1-\mu)$. Averaged over solid angle,
$\langle \Delta \lambda \rangle = 1$. So after $n$ scatterings,
$\langle \lambda_n \rangle = n + \lambda_0$ where $\lambda_0$ is 
the original wavelength (omitted for $n>1$). In the diffusion regime,
the dispersion of photon wavelengths is also propotional to the number
of scatterings. Hence the following relation between the number of 
scatterings or time (in units of scattering time): $\Delta \tau \propto 
\lambda ^2 \propto E^{-2}$. For a finite medium, this 
diffusion process lasts for a number of scatterings $n \propto \tau_T
^{\phantom{T}2}$, thus the maximum width (in wavelegth) of the escaping
photons would be $\Delta \lambda \propto \tau_T $.
In this case, however, the precise law between 
the pulse width and the energy depends on the photon source distribution 
within the scattering medium, as it will be exhibited in more  detail in
the next section. Nonetheless, the qualitative correlation between the 
energy and duration of pulses will always be as that outlined above.

In the following, we make the above heuristic arguments more concrete
by presenting well formulated solutions of the problems outlined
above. 

\bigskip

\ni {\it II a. Relativistic Electron Cooling}

\medskip

Consider the simplest case of time-dependent synchrotron radiation
by a population of non-thermal electrons: Relativistic
electrons of Lorentz factor \g, with a power law distribution of index 
\G~ i.e. $Q_e(\g)= q \g^{-\G}$, are injected in a volume of tangled 
magnetic field $B$ at
an instant $t= t_0$ and are left to cool by emission of synchrotron (on
inverse Compton radiation). The equation which governs the evolution 
of the electron distribution function $N(\g,t)$ in time is

$$\eqalign{{\partial N \over \partial t} &= - {\partial \over \partial \g} 
\left [ \langle \dot \g \rangle N(\g,t) \right] + Q_e(\g) \delta (t -
t_0) \cr
&= {\partial \over \partial \g} \left [ \beta \g^2 N(\g,t) \right] + 
Q_e(\g) \delta (t - t_0) \cr} \eqno (1)$$

\ni where $\langle \dot \g \rangle = - \beta \g^2$ is the mean electron loss rate
and  $\beta = (4/3) (B^2 /8 \pi)(\sigma_T/m_e c^2)c$, while $Q_e(\g)$,
as given above, 
represents the injection of relativistic electron population. This equation 
can be solved by the standard technique of converting it into an ordinary
differential equation in $t$ and integrating it along the characteristic
curves $\g = \g (t)$. Thus, multiplying through by $\g^2$ one obtains

 $$\eqalign{{\partial [\gamma ^2 N ]\over \partial t} - \beta \gamma^2
{\partial \over \partial \gamma} \left[  \g^2 N(\g,t) \right] &= 
\g^2 Q_e(\gamma) \delta (t - t_0) \cr
{d (\g^2 N) \over dt} &= \g^2 Q_e(\g ) \delta (t - t_0) \cr } \eqno(1a)$$

\ni with the last step becoming obvious on identifying $\partial \g / 
\partial t$ with $- \beta \g^2$. The total differential of equation (1a)
can be integrated along the integral curves of the electron energy
\g; these are given by the solution of $\partial \g / \partial t = - \beta \g^2$, 
namely $\g = \g_0 /(1 + \g _0 \beta t)$, where $\g_0 $ is the initial
electron energy. Assuming $Q_e(\g) = q \g^{-\G}$, the resulting form
of the electron distribution function is 

$$N(\g,t) = \cases{
q \g^{-\G} (1 - \beta \g t)^{\G-2} & for $\g < 1 / \beta t $
\cr
0  & for $\g > 1 / \beta t$ \cr}~. \eqno(2)$$

%$$N(\g,t) =
%\eqalign { 
% q \g^{-\G} (1 - \beta \g t)^{\G-2} & ~~ {\rm for}~~ \g < {1 \over \beta t} 
%\cr
% 0 ~~~~~~~~~~~~~~~~~~~ & ~~ {\rm for}~~ \g > {1 \over \beta t} \cr}
%\eqno(1)$$

Given  the electron distribution function $N(\g,t)$, the emitted spectrum 
can be calculated by convolving $N(\g,t)$ with the synchrotron emissivity 
$\epsilon(\g, \nu)$ (see e.g. Rybicki \& Lightman 1979). However, the form
of the emerging spectrum can be easily calculated using the 
$\delta$-function approximation to the single electron emissivity, i.e. 
$\epsilon(\g, \nu) \propto \delta (\nu - \g^2 \nu_c)$, where $\nu_c \sim 
4\times 10^6 B(G)$ Hz is the cyclotron frequency. Assuming, further, 
that the injected distribution function has a minimum Lorentz factor 
$\g_m(t)$ the resulting spectrum has the form

%$$
%\eqalign{   \nu^{1/3} ~~~~~~~~~~~~~~~~~~~~~~~~~~~~~~~ &{\rm for}~~ 
%                              \nu <\nu_m(t) = \g_m(t)^2 \nu_c  \cr
%F_{\nu} \propto  {1 \over 2} \left( {\nu \over \nu_c} \right)^{
%-{\G - 1 \over 2}} \left [ 1 - \left( {\nu \over \nu_c} \right)^{1/2}
%\beta t \right]^{\G - 2}~~ &{\rm for}~~ \nu > \nu_m(t) \cr
%  0~~~~~~~~~~~~~~~~~~~~~~~~~~~~~~~ & {\rm for}~~ \nu > {\nu_c 
%                       \over \beta^2 t^2} \cr} \eqno(3) $$

$$F_{\nu} \propto \cases{ \nu^{1/3}  &{\rm for}~~ 
                              $\nu <\nu_m(t) = \g_m(t)^2 \nu_c$ \cr
{1 \over 2} ( \nu / \nu_c )^{-{\G - 1 \over 2}} 
\left [ 1 - ( \nu / \nu_c )^{1/2}
\beta t \right]^{\G - 2}~~ &{\rm for}~~ $\nu > \nu_m(t)$ 
\cr
0  & {\rm for}~~ $\nu > \nu_c / \beta^2 t^2$  \cr} \eqno(3) $$

\ni where  $\nu_m(t)$ is the synchrotron frequency corresponding to the
lowest energy electrons of the injected electron Lorentz factor $\g_m$
(which is also a function of time), and $\nu_c$ is the cyclotron frequency 
corresponding to magnetic field $B$.

Figure 1 shows the time dependent spectra obtained by integrating the
distribution function of Eq. (2) over the exact expression of the 
synchrotron emissivity, with an injection spectrum $\propto \g^{-2}$
between Lorentz factors $\g_m = 10^6$ and $\g_M = 10^9$. 
These specific values (especially that of $\g_m$) were chosen to assure that
the low energy turnover of the spectrum occurs at $\nu_m \sim 10^{20}$ Hz
(= 414 keV), as indicated by the GRB spectra (Schaeffer \ea 1994), 
and that the duration of the burst be
$\tau_b \sim$ a few seconds, also in agreement with observations. These
considerations then serve to also  determine the value of the
magnetic field associated with this emission to be $B \sim 30$ Gauss. 

If the observed durations are indeed only apparent  rather than  
intrinsic, the result of emission from a plasma in relativistic motion 
toward the observer, as suggested e.g. in the currently popular
relativistic blast wave GRB models (see e.g. M\'esz\'aros \& Rees 1992), 
then both the break frequency $\nu_m$ 
and the characteristic time scale $\Delta \tau$ should be shifted by the
appropriate powers of the bulk Lorentz factor of the flow \G to 
reflect the values at the rest frame of the fluid. In this case,
the intrinsic frequency  of emission, $\nu_m^{\prime}$,
will be shorter from the observed one $\nu_{m,o}$  
by a factor \G, i.e. $\nu_m^{\prime} \simeq
\nu_{m,o}/\Gamma$ while the intrinsic duration $\Delta \tau ^{\prime}$
would be longer than the observed one $\Delta \tau \simeq$ a few sec, 
by a factor $\delta^{-1} = \Gamma(1-\beta_L) \simeq \G$, i.e., 
$\Delta \tau ^{\prime} \simeq \Delta \tau \cdot 
\Gamma$ ($\beta_L$ is the velocity of the fluid, i.e. 
$\G = (1-\beta_L)^{-1/2})$. 
Using the primed (rest frame) values of the duration and
break frequency to determine our values of $B$ and $\g_m$ leads to
the following scalings for the magnetic field $B$ and the value of 
$\g_m$ on  the rest frame of the fluid in terms of the observed 
duration $\Delta \tau($sec) and $\nu_{m,o} = 10^{20} \nu_{20}$ Hz

$$B \simeq 10^{5/3} \Gamma^{-1/3} \nu_{20}^{-1/3} 
\Delta \tau^{-2/3} ({\rm sec}) ~~{\rm Gauss}$$

and 

$$\gamma_m \simeq 10^{17/3} \Gamma^{-1/3} \nu_{20}^{1/3} 
\Delta \tau^{1/3} ({\rm sec})$$
 
So while our calculations were performed with 
a field value of $B \sim 30 $ Gauss and $\g_m \simeq 10^6$
the values of these parameters in the fluid rest frame
would depend on the value of the bulk Lorentz factor \G~ by
the relations given above.

Figure 2 shows the synchrotron emission as a function of time  at a 
set of frequencies, spaced logarithmically, with each curve corresponding 
to a frequency $\times$ 10
smaller than the previous one. It is apparent in this figure that the 
burst duration is inversely proportional to the square root of the 
radiation frequency, in accordance with the relation reported by 
Fenimore \ea (1995). It is also apparent in this figure that, 
for the particular form of electron injection considered in this example
(i.e. impulsive injection), there is no time lag in peak emission between 
frequencies in the range $\nu > \g_m(t_0)^2 \nu_c $, i.e. for
frequencies higher than the synchrotron frequency of the lowest energy 
electrons at the time of injection. Time lags in peak emission  develop 
only for frequencies smaller than that above, resulting from the 
``cooling" of the lowest energy electrons to energies $\g_m(t) < \g_m(t_0)$.

\bigskip

\ni {\it II b. Energy Downgrading by Electron Scattering}

\medskip

Consider alternatively the injection of photons of energy $E \sim
m_ec^2$ near the center of a cold ($T_e \ll E$), spherical (for 
simplicity) electron cloud of Thomson depth $\tau_T \sim 5$. 
As the photons random walk out of the
cloud, they also lose energy in collisions with the ambient electrons. A
monoenergetic photon pulse at the center of the cloud spreads,
therefore, both in energy and in time. The duration of the pulse for
photons of a given energy can be computed from the following
considerations:

At a given scattering event the fractional change of the photon
energy is 

$${\Delta \nu \over \nu} = - {(1 - \mu) h \nu \over m_e c^2}
\eqno(4)$$

\ni or, expressed in terms of the photon wavelength $\l = m_e c^2 / h \nu$,

$$\Delta \l = 1 - \mu ~, \eqno(6)$$

\ni where $\mu$ is the cosine of the angle between the incident and
scattered photon momenta. 

Let $\phi(\mu) = (3/16 \pi) (1 + \mu^2)$ denote the phase function of
the Thompson cross section, such that $\int \phi(\mu) d \Omega =1$;
then the average change in the photon wavelength $\langle \Delta \l
\rangle$ and the dispersion in the photon energies $\sigma_{\l}^2$
per scattering are

$$m_{\l} = \langle \Delta \l \rangle = \int \phi(\mu) (1 - \mu) d
\Omega = 1 \eqno(7)$$

\ni and

$$\eqalign{ \sigma_{\l}^2 & = \langle (\Delta \l)^2 \rangle - \langle 
\Delta \l \rangle ^2 = \langle (\Delta \l)^2 \rangle - 1 \cr
& = \int \phi(\mu) (1 - \mu)^2 d \Omega - 1 = {3 \over 8}
\int_{-1}^{1} (1+ \mu^2)(1-\mu)^2 d \mu = {2 \over 5} \cr } \eqno(8)$$

The distribution (probability) of photons over number of scatterings
$n$ with a mean number of scatterings $u$ is given by a Poisson 
distribution, $P_u(n)$,  which is the limit of binomial distribution for the 
case of rare  events. For the Poisson distribution the values of the
mean $m_n$ and dispersion  $\sigma_n^2 $ are equal, i.e. $\sigma_n^2 = m_n = u$.  

Now, the photon distribution in wavelength \ls for a given mean  number
of scatterings $u$ (or alternatively after time $u$ measured in units
of the scattering time $ \tau = 1/\sigma_Tn_e c$) is given by the
convolution

$$J_u(\l) = \sum_{n=0}^{\infty} \Psi_n(\l) P_u(n) \eqno(10)$$

\ni where $\Psi_n(\l)$ is the photon distribution after $n$
scatterings, with mean wavelength $\langle \l_n \rangle = n + \l_0
~(\l_0$ is the wavelength of the initial photon) and dispersion
$\sigma_{\l}^2(n) = n \cdot \sigma_{\l}^2 = (2/5) n$ and $P_u(n)$ is 
the probability of $n$ scatterings within time $u$.

The mean of this distribution is 

$$\eqalign{ m_{\l}^u & = \int_{\l_0}^{\infty} J_u(\l) d \l =
  \sum_{n=0}^{\infty}\left( \int_{\l_0}^{\infty} \l \Psi_n(\l) d \l
\right) P_u(n) \cr
  & = \sum_{n=0}^{\infty} (n + \l_0)P_u(n) = u + \l_0 \cr} \eqno(11)$$

\ni and the dispersion

$$\eqalign{ \sigma_{\l}^2(u) & = \int_{\l_0}^{\infty} (\l - m_{\l}^u)^2
J_u(\l) d \l \cr
 & = \int_{\l_0}^{\infty} \left[(\l - m^n_{\l})^2 + 2
(\l-m_{\l}^n)(m_{\l}^n - m_{\l}^u) + (m_{\l}^n - m_{\l}^u)^2 \right]
d \l \, P_u(n) \cr
 & = \sum_{n=0}^{\infty} \left[ {2 \over 5} n + 0 + (n-u)^2 \right]
P_u(n) \cr
 & = {2 \over 5} u + u  = { 7 \over 5} u \cr } \eqno(12)$$

\ni Therefore the photon spectrum  after time $u$ will have the form

$$J_u(\l) = {1 \over \sqrt{2 \pi \sigma_{\l}^2(u)}} exp \left[- 
{(\l - m_{\l}^u)^2 \over 2 \sigma_{\l}^2(u)} \right]~, \eqno(13)$$

\ni which represents a gaussian of width $u$ about $\l - \l_0$. In fact, 
the above relation can be viewed both as the photon spectrum escaping after
time $u$ or alternatively as the time history for photons of a given
wavelength \l. One can use the above expression to obtain the dispersion 
in escaping time for photons of a given energy. This is simply given by
the width of the gaussian in equation (13), which, considering the 
definitions of $\sigma_{\l}^2(u)$ and $m_{\l}^u$, yields 
$\l - \l_0 \simeq 2 \sqrt{2 u}$, i.e. that $\Delta \tau \propto E^{-2}$,
as suggested by our heuristic arguments.

The emergent spectrum of the photons escaping over the duration of
the burst is given by the convolution 

$$I(\l) = \int_0^{\infty} J_u(\l) P(u) d u  \eqno(14)$$

\ni where $P(u)$ is the probability of a photon to escape after $u$
scatterings. As a concrete example, we consider a photon source 
distribution according to the first eigenfunction of the diffusion 
operator; with this assumption, the function $P(u)$ has the 
form 

$$P(u) = \beta e^{-\beta u} \eqno(15)$$

\ni with $\beta = \pi^2/(3(\tau_0+2/3)^2) $ being the eigenvalue 
of the same operator. Then the integral of Eq. (14) above is analytic 
yielding

$$I(\l) = {\beta \l^2 \over (1+ {7 \over 5} \beta)^{1/4} 
(1+ {7 \over 10} \beta)^{1/10}} exp[-(\l -\l_0) \beta] \eqno(16) $$

\ni which indicates that the spectrum is a power law in energy of index
$-2$ with an exponential turn--over at $\l \sim 1/\beta$.

The analytic results of the above discussion should be viewed only as an 
illustrative example to indicate the relationship between the 
photon energy $E$ and the duration of a photon pulse $\Delta \tau$. 
It is also apparent from the above discussion that the precise form 
of this relation depends on the distribution of photon sources
within the scattering cloud, the relation derived from equation (13)
correponding to a homogeneous, infinite medium.
Our analytic calculations are also more appropriate for
the case in which the high energy photon source at the center of 
the electron cloud. This ensures that the escaping photons will 
undergo quite a few scatterings before escape and that they will lose a 
significant fraction of their energy in the first few scatterings.
These scatterings reduce their energy to the point where the treatment of 
radiative transfer can be replaced by diffusion in energy and space 
(i.e. small energy change per scattering at the Thomson cross section).
Clearly, the random walk of photons situated close to the cloud
boundary, cannot be described by diffusion with small energy change
at each scattering event. A large fraction of the high energy photons
(assuming monoenergetic high energy injection) undergoes only a 
few scatterings before escape but with large energy change per
scattering (Hua \& Titarchuk 1996). Thus the accurate consideration
of the photon transport distributed over a finite plasma cloud
could, conceivably, lead to an energy -- duration 
relation in closer agreement to the results of Fenimore \ea (1995).

Because our analytic caclulations are appropriate only when the 
source of photons is located at the center of the cloud,
we have extended our model to make it more realistic at the 
expence of introducing an additional parameter. 
The extended model consists of two concentric spheres;
it is assumed that cold electrons are uniformly distributed 
over the entire volume of the larger sphere of radius $R_0$.
The smaller sphere, of radius $R_s$, represents the volume within
which the high energy photon sources are located and the photon 
injection takes place, uniformly over its entire volume. This model
has an additional parameter namely the ratio of the radii of these 
two spheres $R_s/R_0 = \tau_s/\tau_0$. It is also assumed that the photon 
injection takes place impulsively at $t = 0$, at energies $\sim 
1$ MeV.  For the treatment of this more detailed model we have
used a Monte--Carlo code developed by one of us (XMH). 
The details of this Monte--Carlo code have been discussed elsewhere
and will not be repeated here. The interested reader is referred
to the relevant publications (Hua \& Titarchuk 1995).

In Figure 3 we present the widths in time  associated with photons 
escaping frorm the scattering cloud at several energy bands, 
as obtained by our Monte-Carlo calculation; the energy bands were 
chosen in the same fashion as in the 
analysis of Fenimore \ea (1995). The widths of the pulses 
as a function of the energy are given for three values of 
the ratio $\tau_s/\tau_0$, namely 1.0, 0.5 and 0.1. 
The photon injection was assumed to be monoenergetic at $E_0 = 1$
MeV and the total Thomson depth of the source was taken to be
$\tau_0 = 10$. One can see that the value $\tau_s / \tau_0 = 0.5$ 
provides good fit to the relation found by Fenimore \ea (1995), 
indicated by the straight solid line. 

The resulting photon spectrum, in $\nu F_{\nu}$ units, 
accumulated over the entire
burst is given in Figure 4, with the spectra escaping at specific
times given by the individual curves. The times are spaced equally in
logarithmic intervals. The time-accumulated spectrum is consistent
with that of GRBs in that it peaks at roughly 1 MeV.
Clearly, injection of photons at higher energies and/or at larger
Thomson depths could produce spectra which would simulate the
observed ones more  accurately.
Finally, in figure 5 we present the rates of escaping photons 
as a function of time, for photons of a given energy range. 
The energy ranges were chosen to be identical to those used by 
Fenimore \ea (1995) to derive the energy -- duration relation.

\bigskip

\ce{\bf III. Conclusions, Discussion}

\medskip

We have presented above two models which can account for the recently
discovered  correlation  of the GRB duration (or the duration of 
subpulses within a given burst) as a function of 
the photon energy. Both models appear to be able to reproduce  
the observed power law dependence of $\Delta \tau$ on $E$ 
given in Fenimore et al. (1995), as well as the general form of the 
observed GRB spectra.
Considering in greater detail the comparison of our model spectra with 
those of GRBs (Schaeffer \ea 1994), the model spectra are generally 
consistent with those observed in that their 
luminosity (i.e. their $\nu F_{\nu}$ distribution) peaks at $\sim 10^{20}$ Hz, 
while exhibiting a power law low--frequency dependence, in agreement with 
observations. 

The slope of this low--energy power law is very well determined
for the synchrotron cooling models, being proportional to $E^{4/3}$, the
functional form of the single electron synchrotron emissivity. 
Interestingly, there exist several GRBs (e.g. GRB 910503 and GRB 910601)
whose low frequency spectra are in agreement with this interpretation.
In particular, the spectrum of GRB 910601 (Share \ea 1994), seems to be fitted
very well with the single electron synchrotron emissivity 
over its entire energy range, suggesting the synchrotron origin of 
the radiation associated with this specific burst.
However, most of the spectra in Schaeffer \ea (1994) 
require low--frequency power law fits of significantly different slopes,
typically $\nu F_{\nu} \propto  E^{1/2}$. Clearly, the simple interpretation
of synchrotron emission by a uniform, optically thin, impulsively 
injected electron population, such as described above,
would be inappropiate for this type of bursts. However, more 
complicated models which relax the above assumptions could produce spectra
agreement with observation. For example, continuous (rather than 
impulsive) injection of the electron population over time scales 
long compared to the loss time scale of the electrons with Lorentz
factor $\gamma_m$ would lead to spectra in agreement with this 
form. 

On the other hand, the downscattering model could accommodate 
spectra of the form $\nu F_{\nu} \propto  E^{1/2}$, since the 
precise slope of the low energy power law depends on the distribution
of the high energy photon sources within the cold cloud considered. 
These slopes are consistent with those of the time averaged 
spectrum of the  downscattering model given in figure 4. 
In further support of this point, one should bear in mind that 
the spectrum of figure 4 represents the result of a $\delta$--function photon
injection at the highest energy bin, hence the
upward turn near $\simeq 1$ MeV in figure 4. Injection of a broader type 
of photon spectra or larger value for $\tau_0$ would result in smoother
spectra,  conforming more to the observation. 

In assessing the utility of the above models,
one could  claim that both may be of relevance in modeling the 
spectro -- temporal evolution of 
GRBs. The synchrotron cooling model fares a little better in 
reproducing the duration -- energy correlation in that it
requires fewer parameters to do so than the downscattering model. 
On the other hand, the downscattering model, because of its additional 
freedom, appears to be able to provide more detailed fits to a larger 
number of GRB spectra. Discriminating between these models 
requires  a more detailed combination of spectral and temporal analysis 
of individual bursts rather than the use of their average properties.
Nonetheless, the form of the observed correlations along with their 
present model interpretation clearly indicate that, whatever 
their precise mechanism, the available energy of GRBs is 
originally released into the highest energy particles and their
evolution is determined by its subsequent cascade to the lower 
energy ones. 

Despite the general agreement of the computed spectra of 
both models to those observed and their overall agreement 
to the duration -- photon energy relation it is hard, at the present 
level of analysis, to draw any further conclusions concerning the physical 
mechanism underlying the emission of radiation in GRBs. This is due to
the absence of an independent estimate of the parameters
which set the scales of the time and the energy of peak emission in 
the $\nu F_{\nu}$ spectra, in either model. In the 
non--thermal model, the time is set by the magnitude of the magnetic 
field, or rather the combination $B^2 \g$ (or by $\rho_{rad} \g$ if 
inverse Compton is the dominant loss mechansim, and the bulk motion
Lorentz factor \G~ if relativistic expansion is of relevance),
while in the downscattering model it is set by the density 
of the electron cloud through which the \g--rays propagate, both of
which are indeterminate. The energy at which the $\nu F_{\nu}$ spectra
peak could be used to provide some additional information on the 
radiative processes associated with GRBs. However, such information
is independent, at this level of analysis, from that 
associated with the GRB time scales. In the non--thermal models, 
the energy of peak emission 
is set by the combination $\g_m^2 B$, while in the downscattering
models it is an independent parameter, defined by the injection 
of primary photons.

Figures 2 and 5 exhibit 
an additional aspect of the spectro -- temporal correlations of the
GRB time profiles which has been touched upon only briefly so far, 
namely that of the time lags in peak emission among the various 
energy channels. This issue, only a peripheral one in this note, 
could be further developed 
both theoretically and observationally and might provide
additional information on the nature of the GRB emission process,
or at least a sufficiently strong criterion to reject one or both
of the above models.

Figure 5, indicates that  the model of  impulsive high energy
photon injection leads invariably to lags in the peak (highest 
intensity) emission between the various energy channels, 
in the sense that the lower energy emission always lags behind 
the high energy one. Due to the non$-$dimensional character of the
time axis of this figure, the lags have to be compared to the dispersion
(FWHM) of the appropriate energy photons. It is apparent in the 
figure 5 the lag is smaller
than the dispersion (FWHM) of the burst, in qualitative agreement
with the results of fig. 12 of Norris \ea (1996).
While the same data indicate generally lags significantly smaller
than the observed FWHM of the pulses, we believe that more 
detailed study is required to decide the quantitative agreement
of the model with the data and possibly exclude this model 
on the basis of this issue.

In the relativistic electron cooling model, on 
the other hand, the presence of such lags is not ubiquitous,
even for an impulsive injection of the electron population. 
As shown in figure 2, lags  develop only for frequencies 
emitted by electrons with Lorentz factors lower than $\g_m(t_0)$,
as discussed earlier. Indeed, as seen in figure 2 there is little 
lag in the peak emission between frequencies $\gsim 10^{20.5}$ Hz, 
while there are progressively increasing lags between the peak emission
of lower frequencies. Since the time axis of figure 2 is absolute 
(i.e. in seconds) one can compare these lags direclty to those 
discussed in  Norris et al. (1996).  In doing so, 
one should bear in mind that the lowest channel
of the above reference corresponds to curve $f$ of this figure so that 
no lags longer than 0.5 sec should be expected for the chosen 
values of the parameters, in rough agreement with Norris \ea (1996).

The issue of lags in peak emission within both models can be further 
complicated by the relation of the injection to the loss 
time scales in cases in which the latter is not impulsive. 
A further indication of the complex nature of this issue is the
fact that, as noted by Norris \ea (1996), the lags in peak emission 
depend not only the energy band of observation but also  the asymmetry
in the burst time profiles, an issue not discussed here.
We prefer not to delve into these issues in this note but concentrate
instead on the relation between the observed photon energy and the
burst duration, since that is, in both our models,
property of the Green's function of the associated problem and 
hence independent of the details of injection.
Some of the aspects of their findings could be accounted by the models 
presented above. However, since the entire burst time profile seems 
to be of importance in this issue, we will defer addressing it 
to future more detailed models of the spectro -- temporal evolution 
of GRBs. The systematics provided by Norris \ea (1996) indicate 
that this issue is an important one and in need of further
theoretical exploration.   

Finally, we would like to point out that neither  model
provides an answer to why GRBs emit most of their luminosity 
in the \gr s rather than at other wavelengths.
The downscattering model treats the injection of primary photons
as an initial condition and it is hence beyond the scope of 
any further discussion.
The synchrotron emission model actually does fare a little better 
in this respect, since \g --ray emission is the 
natural outcome of electron acceleration in shocks: 
It can be easily shown that for shock acceleration of electrons
limited by synchrotron losses, 
the maximum Lorentz factor is of order $\g^2 \sim (e/fB)(1/ 
\sigma_T)$ and the frequency of the emitted radiation $E_{\g} 
\sim (1/8f)(c / r_0) \simeq 2 \times 10^{22} {\rm Hz}  \sim (24/f)$
MeV ($r_0$ is the classical electron radius and $f$ is a measure of 
the electron free path in units of the Larmor radius), which has 
the interesting property of being independent of the magnetic field $B$.
If this is indeed the reason that GRBs emit mainly in the \g-ray band,
it would not leave much room for the relativistic blast wave
models (M\'esz\'aros and Rees 1992), since the latter postulate 
an additional relativistic boost of the produced photons with bulk
Lorentz factors $\G \sim 10^3$ in order that they are detected
in the \g--ray band. 

The models presented above, despite their ability to provide good 
fits to the energy -- duration relation of GRBs, they cannot, 
at the present level of analysis, provide an immediate, decisive 
answer to the question of the nature
of GRBs. Nonetheless,  the energy -- duration relation reported by 
Fenimore \ea (1995) which represents the tightest
correlation associated with GRBs to date, warrants their use 
in more detailed spetro -- temporal analyses of GRB time profiles, 
which may set the ground work for uncovering the nature of GRBs.

We would like to acknowledge useful discussions with M. Baring, 
J. Norris, J. Contopoulos as well as useful, meaningful correspondence 
with the referee of our paper.

\bigskip

\ce{\bf References}

\bigskip

\ref Fenimore, E.E.,  \eaa, 1995, ApJ, 448, L101

\ref Fishman, G., \eaa, 1992, in Gamma--Ray Bursts: Huntsville, 1991,
ed.W. S. Paciesas \& G. J. Fishman (New York: AIP), 13

\ref Hua, X. M. \& Titarchuk, L. G., 1995, ApJ,449, 188

\ref Hua, X. M. \& Titarchuk, L. G., 1995, ApJ,469, 280

%\ref Ford, L.A., \eaa,  1995,  ApJ, 439, 307

%\ref Kouveliotou, C., et al.  1993,  ApJ, 413, L101

\ref Link, B., Epstein, R.I., \& Priedhorsky, W.C.  1993, ApJ, 408, L81

\ref Meegan, C. A. \eaa, 1992, Nature, 355, 143

\ref M\'esz\'aros, P. and Rees, M.J., 1992, ApJ, 397, 570

%\ref Nemiroff, R.J., et al., 1994, Ap.J., 423, 432

\ref Norris, J.P., \ea  1987, Adv.  Space Res., 6,  19.

\ref Norris, J.P., et al.  1994, in Gamma-Ray Bursts - Huntsville, AL
1993, AIP 307, p.  172

%\ref Norris, J.P.  1995, Ap Space Sci, 231, 95

\ref Norris, J.P., \ea  1996,  ApJ,  459, 393

\ref Paczynski, B., 1992, in CGRO St. Luis conf., eds. M. Friedlander, N. Gehrels, 
D. J. Macomb (New York: AIP), 981

\ref Rybicki, G. B.  \& Lightman, A. P., 1979, Radiative Processes in Astrophysics,
(New York: Wiley)

\ref Schaeffer, B. E. \ea, 1994, ApJS, 92, 285

\ref Share, G. et al., 1994, in Gamma-Ray Bursts - Huntsville, AL 1993, AIP 307, p. 283

\ref Sunyaev, R. A. and Titarchuk, L. G., 1980, A\&A, 86, 121

\eject

\ce{\bf Figure Captions}

\bigskip

\bigskip

 Fig. 1. -The evolution of the spectrum with time for the
cooling synchrotron emitting relativistic electron model. The times
corresponding to each curve are: (a) $10^{-3}$ s, (b) $10^{-2}$ s, 
(c) $10^{-1}$ s, (d) 1 s, (e) 10 s. The $B$-field is assumed to be
30 Gauss, the maximum and minimum electron Lorentz factors at 
injection were taken to be respectively $10^9$ and $10^6$ respectively.

\bigskip

Fig. 2.-The synchrotron emission at individual frequencies 
as a function of time, for the relativistic cooling electron model. The 
frequencies corresponding to the various curves are: (a) $10^{24}$ Hz,
(b) $10^{23}$ Hz, (c) $10^{22}$ Hz, (d) $10^{21}$ Hz, (e) $10^{20}$ Hz,
(f) $10^{19}$ Hz, (g) $10^{18}$ Hz, (h) $10^{17}$ Hz.

\bigskip

Fig. 3.-The pulse width -  energy relation for the
downscattering model for different values of the $\tau_s/\tau_0$
ratio.

\bigskip

Fig. 4. -The time-accumulated spectrum, for the downscattering
model described in the text.

\bigskip

Fig. 5. -The time evolution of emission for various energy
channels, for the downscattering model.

\eject

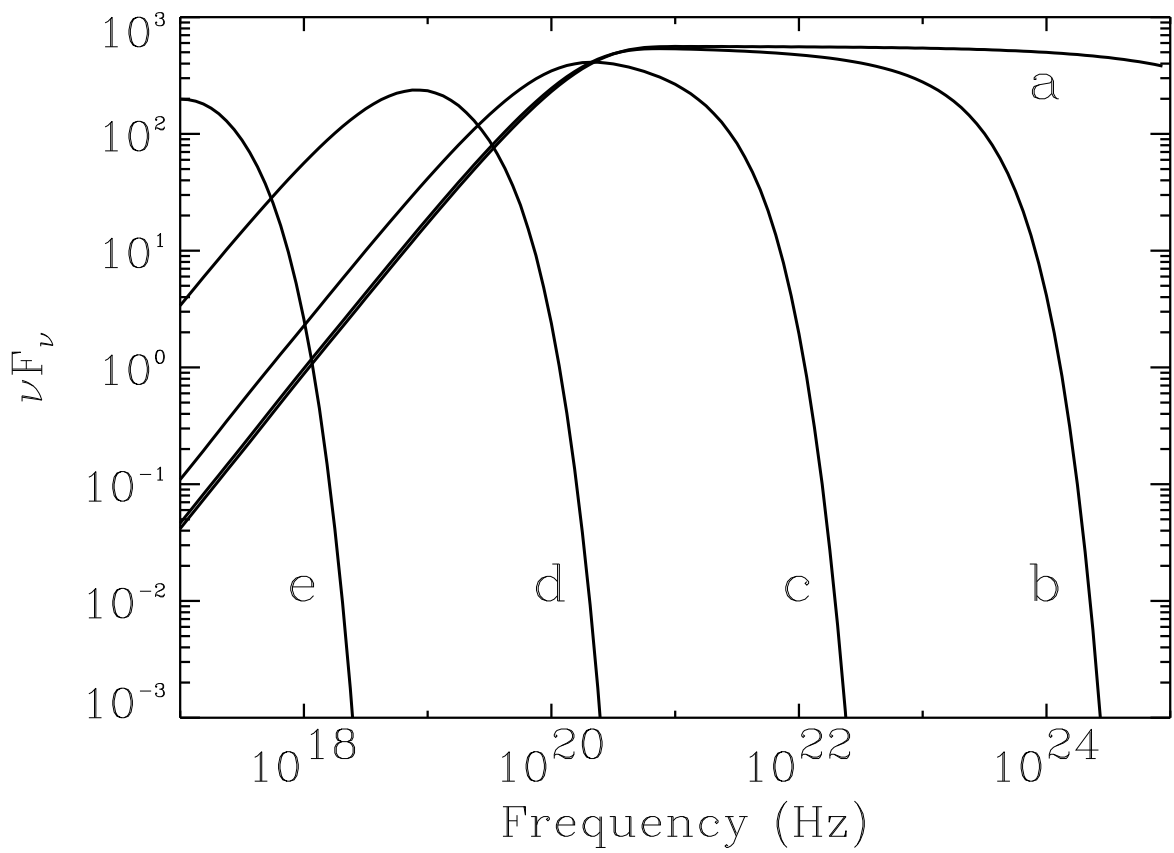
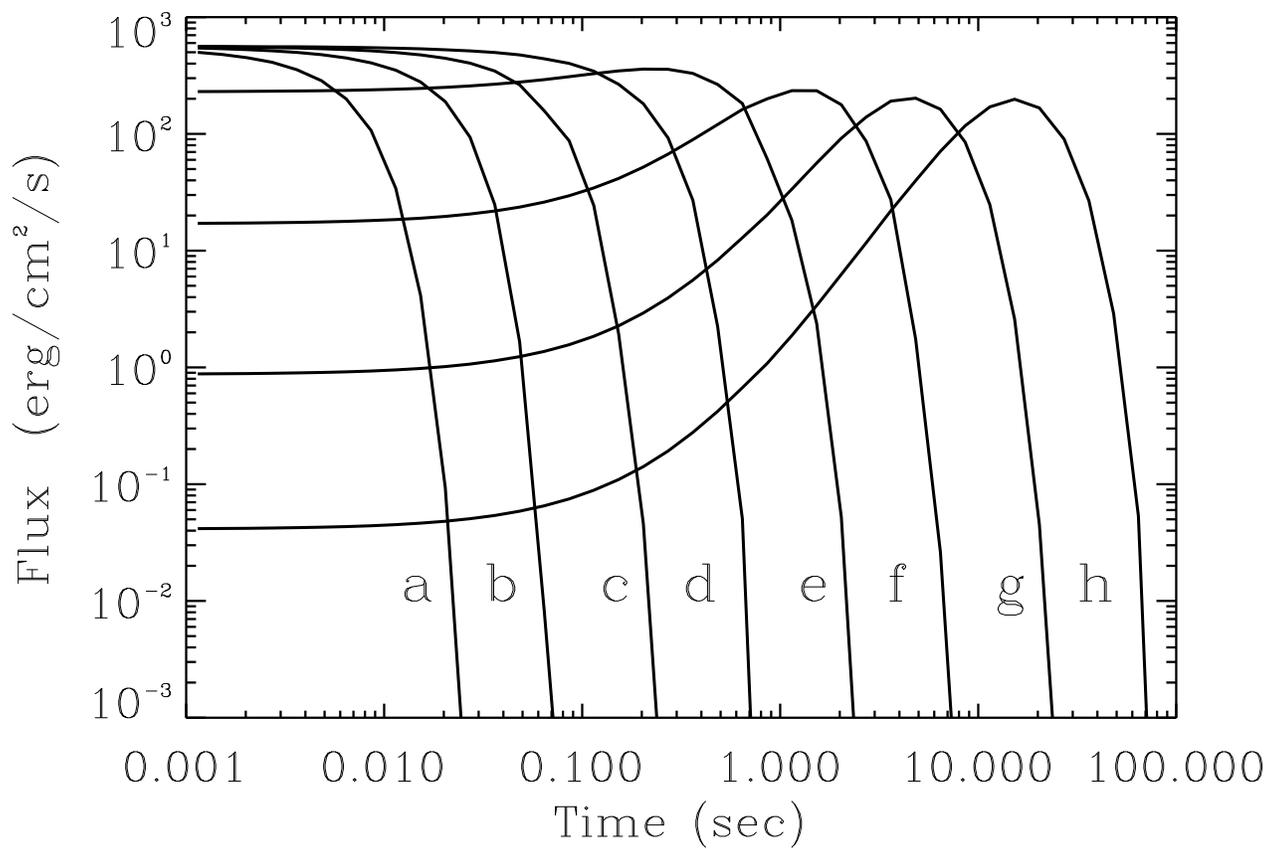
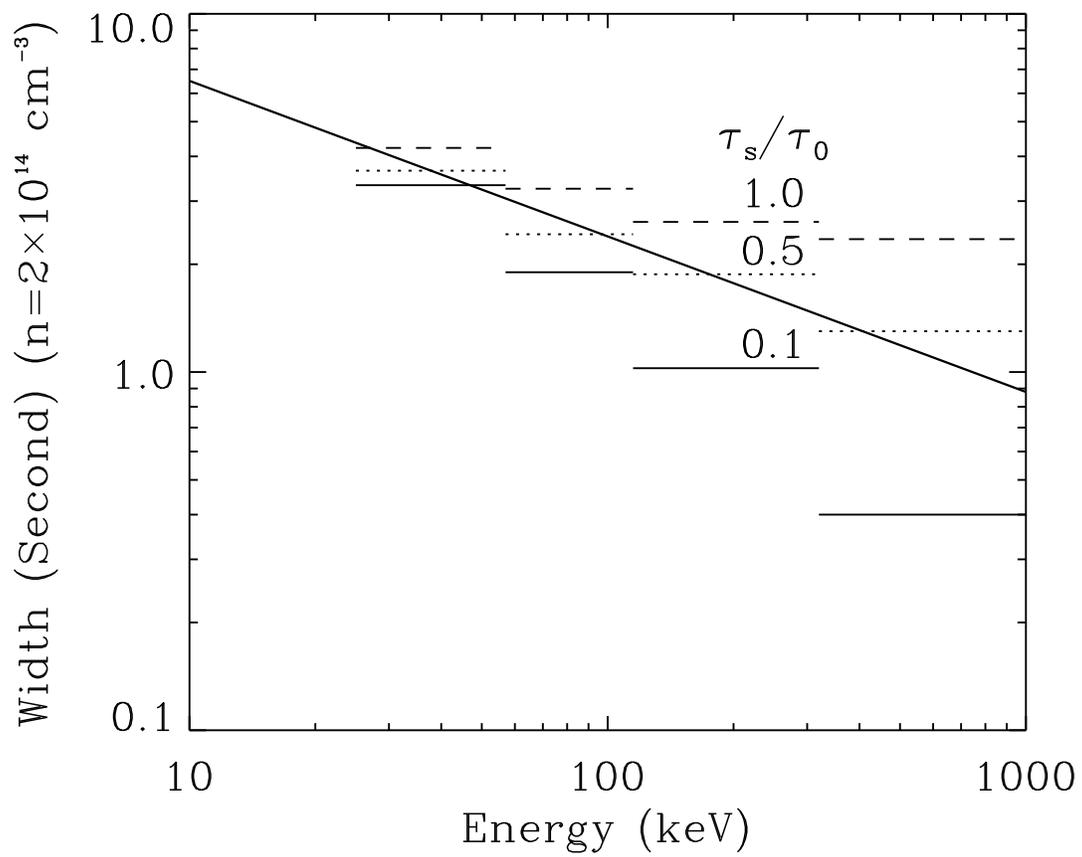
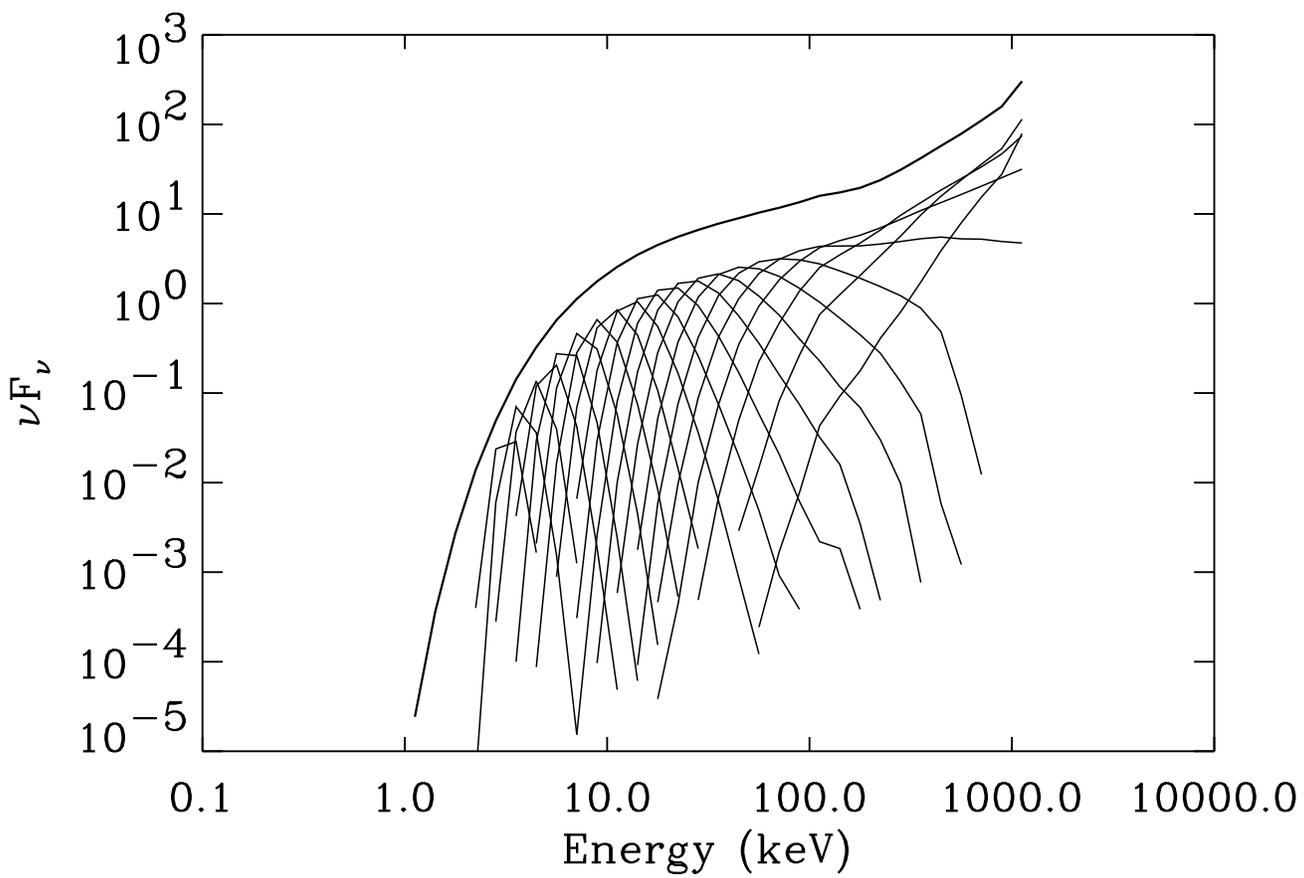
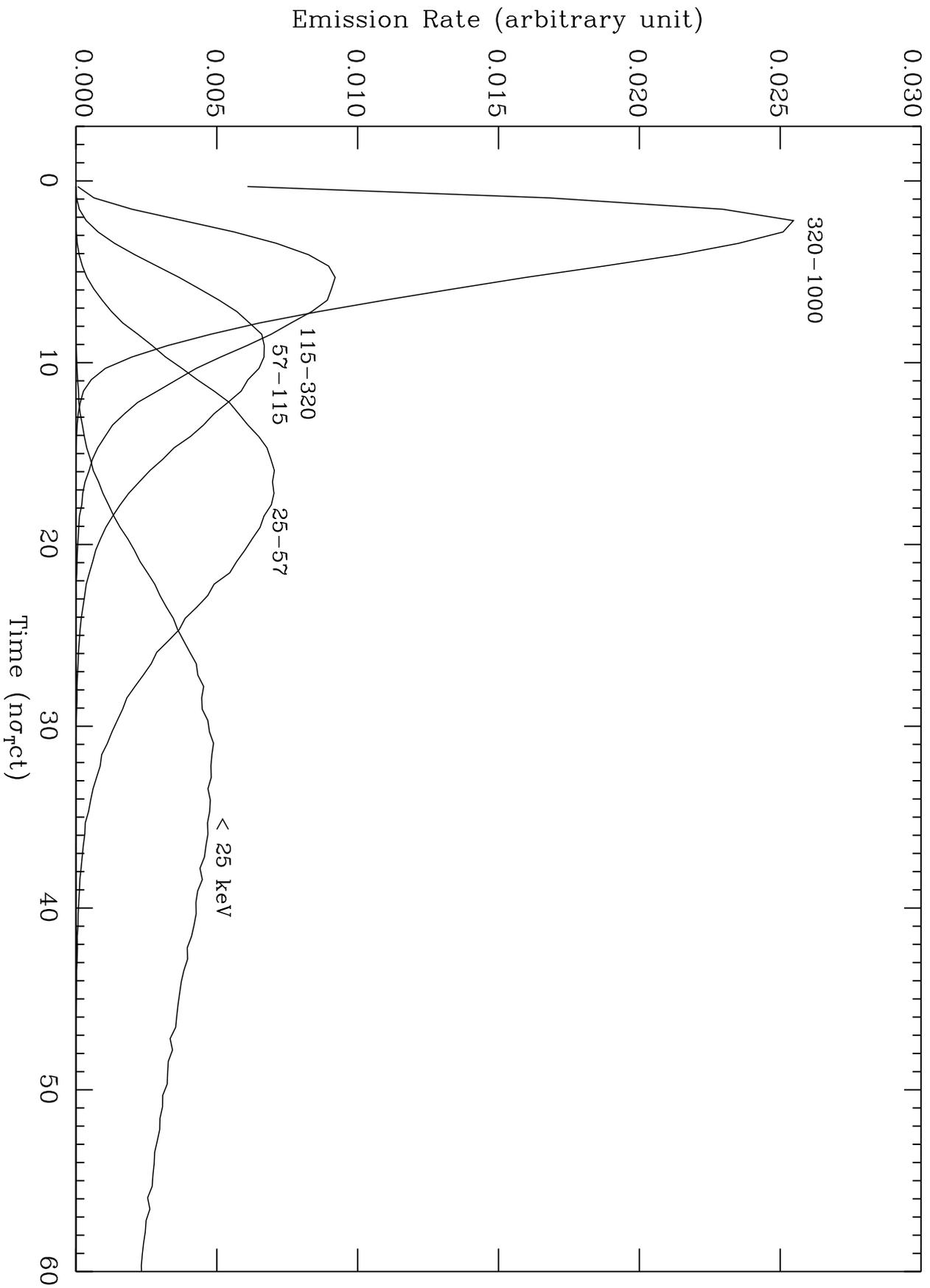
\end